\documentclass[prb,aps,twocolumn,superscriptaddress,showpacs,footinbib,longbibliography]{revtex4-1}
\usepackage[colorlinks=true,linkcolor=black, citecolor=blue, urlcolor=blue, 
unicode=true,breaklinks]{hyperref}

\usepackage{graphicx}
\usepackage{epstopdf}
\usepackage{amsmath}

\usepackage{dsfont} 

\usepackage{pgfplots}
\usepackage{cleveref}

\def\doubleunderline#1{\underline{\underline{#1}}}

\begin{document}

\title{Anisotropic excitonic magnetism from discrete $\mathrm{C}_{4}$ symmetry in CeRhIn$_{5}$}

\author{D. J. Brener}
\affiliation{The Higgs Centre for Theoretical Physics, University of Edinburgh, Edinburgh EH9 3JZ, UK}
\author{I. Rodriguez Mallo}
\affiliation{School of Physics and Astronomy, University of Edinburgh, Edinburgh EH9 3JZ, UK}
\author{H. Lane}
\affiliation{School of Physics and Astronomy, University of St. Andrews, North Haugh, St. Andrews, Fife, KY16 9SS, UK}
\author{J. A. Rodriguez-Rivera}
\affiliation{NIST Center for Neutron Research, National Institute of Standards and Technology, 100 Bureau Dr., Gaithersburg, MD 20899, USA}
\affiliation{Department of Materials Science, University of Maryland, College Park, MD  20742, USA}
\author{K. Schmalzl}
\affiliation{Forschungszentrum Juelich GmbH, Juelich Centre for Neutron Science at ILL, 71 avenue des Martyrs, 38000 Grenoble, France}
\author{M. Songvilay}
\affiliation{Universit\'e Grenoble Alpes, CNRS, Institut Néel, 38042 Grenoble, France}
\author{K. Guratinder}
\affiliation{School of Physics and Astronomy, University of Edinburgh, Edinburgh EH9 3JZ, UK}
\author{C. Petrovic}
\affiliation{Department of Physics, Brookhaven National Laboratory, Upton, New York, 11973, USA}
\author{C. Stock}
\affiliation{School of Physics and Astronomy, University of Edinburgh, Edinburgh EH9 3JZ, UK}

\date{\today}

\begin{abstract}

\noindent Anisotropy in strongly correlated materials is a central parameter in determining the electronic ground state and is tuned through the local crystalline electric field.  This is notably the case in the CeCo$_{x}$Rh$_{1-x}$In$_{5}$ system where the ground-state wave function can provide the basis for antiferromagnetism and/or unconventional superconductivity.  We develop a methodology to understand the local magnetic anisotropy and experimentally investigate with neutron spectroscopy applied to antiferromagnetic ($T_{N}$=3.8 K) CeRhIn$_{5}$ which is isostructural to $d$-wave superconducting ($T_{c}$=2.3 K) CeCoIn$_{5}$.  Through diagonalizing the local crystal field Hamiltonian with discrete tetragonal $\mathrm{C}_{4}$ point group symmetry and coupling these states with the Random Phase Approximation (RPA), we find two distinct modes polarized along the crystallographic $c$ and $a-b$ planes, agreeing with experiment.  The anisotropy and bandwidth, underlying the energy scale of these modes, are tuneable with a magnetic field which we use experimentally to separate in energy single and multiparticle excitations thereby demonstrating the instability of excitations polarized within the crystallographic $a-b$ plane in CeRhIn$_{5}$.  We compare this approach to a $S_{eff}={1\over 2}$ parameterizations and argue for the need to extend conventional SU(2) theories of magnetic excitations to utilize the multi-level nature of the underlying crystal-field basis states constrained by the local point-group $\mathrm{C}_{4}$ symmetry.

\end{abstract}


\maketitle

\section{Introduction}
Unconventional superconductivity is often found to coexist or be coupled with localized magnetic moments.~\cite{Stewart84:56,Mathur98:394,Pfleiderer09:81} Noteworthy examples include the cuprate superconductors~\cite{Proust19:10,Shen08:11,Kastner98:70}, iron based pnictide and chalcogenides~\cite{Johnston10:59,Dai15:87}, and nickel-based materials~\cite{Sun23:621,Canfield98:51,Muller01:64}. While cuprate and iron based compounds display large superconducting transition temperatures making them potentially useful for applications, the underlying energy scales in these compounds are large which makes them challenging to experimentally study. The strongly correlated and cooperative phenomena in these materials is fundamentally defined by single-ion energetics which fixes the ground state wavefunction. With the crystalline electric field defining the ground state magnetism in these materials set by $10Dq\sim$ 1 eV~\cite{Haverkort07:99,Larson07:99,Cowley13:88,Kim11:84}, multiple experimental techniques are required to study magnetic and electronics properties of these compounds spanning a large energy scale. 

Heavy fermion materials based on $4f$ elements, while displaying considerably lower superconducting transition temperatures, have much smaller crystalline electric field energetics and are therefore arguably more amenable for laboratory experiments. Illustrating this is CeCoIn$_{5}$ with a superconducting transition temperature of $T_{c}$=2.3 K~\cite{Petrovic01:13}, yet displaying an unconventional $d$-wave order parameter~\cite{Izawa01:87,Aoki04:16} analogous to that observed in cuprate superconductors at considerably higher temperatures.  $4f$ compounds therefore represent important case studies for competing magnetic and electronic phases.  A unique aspect of magnetic $4f$ ions over $3d$ transition metal ions is the anisotropic crystalline electric field environment with an energy scale that is comparable to the magnetic exchange that couples magnetic ions.  We investigate in this paper the consequences of this and also the crystalline anisotropy on the magnetic excitations.  We focus on the N\'eel ordered phase in CeRhIn$_{5}$ which is a precursor and arguably parent to unconventional superconductivity in the Ce$X$In$_{5}$ series of compounds.~\cite{Thompson12:81,Park12:108}

The heavy fermion Ce$X$In$_{5}$ series of materials display an interplay between various phases including density waves, helical and collinear magnetism, and $d$-wave superconductivity. Neutron spectroscopic studies of CeCoIn$_{5}$ report a magnetic resonance peak~\cite{Stock08:100} tied with the formation of $d$-wave superconductivity, illustrating an underlying coupling between magnetism and superconductivity.  We note that similar spin resonances have been reported in multiple heavy fermion compounds which are both magnetic and unconventional superconductors.~\cite{Stockert10:7,Eremin08:101}   This resonance peak in CeCoIn$_{5}$ is highly anisotropic~\cite{Raymond15:115} being polarized along the crystallographic $c$-axis and is energetically doubly degenerate and can be split with an applied magnetic field.~\cite{Stock12:109,Raymond12:109,Panarin09:78}  Whilst much attention has been given to CeCoIn$_{5}$ with its large (for heavy fermions) superconducting transition temperature, CeRhIn$_{5}$ has also been reported to be a superconductor with a much lower transition temperature of $T_{c}\sim$ 100 mK at ambient pressures.~\cite{Zapf01:65,Paglione08:77,Chen06:97}  At higher temperatures of $T_{N}$=3.8 K, CeRhIn$_{5}$ enters a helical magnetic phase.  Superconductivity is enhanced with the application of pressure which also suppresses helical antiferromagnetism.~\cite{Park08:105}  The goal of this paper is to understand the dynamics and magnetic ground state of the parent antiferromagnetic phase with the objective of relating it to the magnetic properties of its superconducting counterparts and in particular superconducting and non-magnetic CeCoIn$_{5}$. To accomplish this we discuss and theoretically analyze the magnetic dynamics of CeRhIn$_{5}$ as a function of applied magnetic field.

\begin{figure}
	\begin{center}
		\includegraphics[width=80mm,trim=2.5cm 1.2cm 4cm 6.5cm,clip=true]{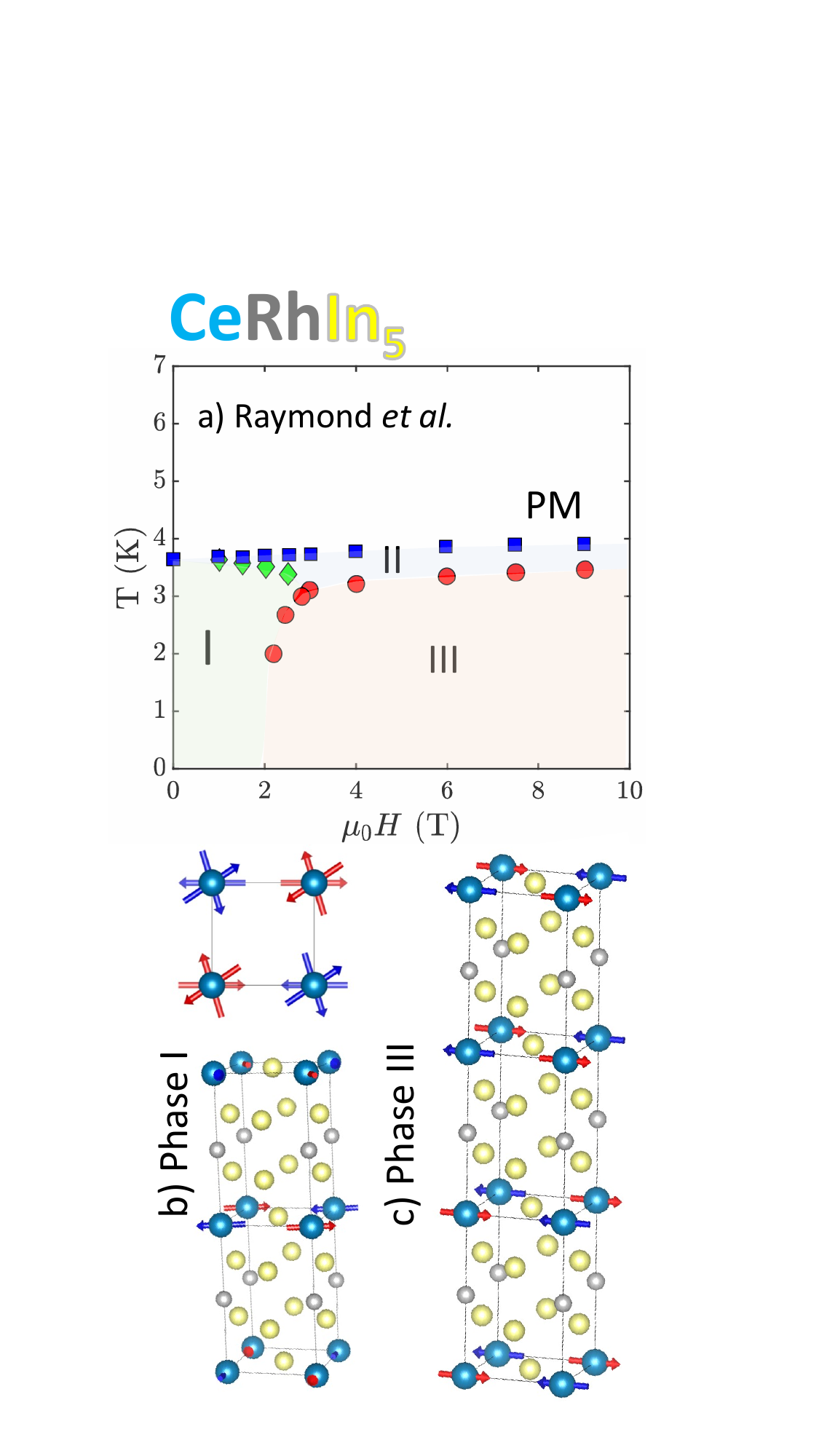}
	\end{center}
	\caption{The low temperature magnetic phases of CeRhIn$_{5}$. $(a)$  The magnetic field-temperature phase diagram measured with neutron diffraction in Ref. \onlinecite{Raymond07:19}.  $(b)$ The zero field helical magnetic structure (Phase-I) defined by a propagation wavevector $\vec{Q}_{0}=({1\over2},{1\over2},0.297)$.  $(c)$ The high magnetic field and low temperature magnetic structure (Phase-III) defined by the wavevector $\vec{Q}_{0}=({1\over2},{1\over2},{1\over4})$.}
	\label{fig:figure1}
\end{figure}

The magnetic field dependent phase diagram of CeRhIn$_{5}$ is illustrated in Fig. \ref{fig:figure1} as mapped out using neutron diffraction in Ref. \onlinecite{Raymond07:19} and displays three distinct phases.  For low temperatures and small magnetic fields (``Phase-I") CeRhIn$_{5}$ magnetism is defined by a helical phase along the crystallographic $c$-axis with the periodicity defined by the wavevector $\vec{Q}_{0}=({1\over2}, {1\over2}, 0.297)$.  Given that it is difficult to distinguish between a spin-density wave (where the magnitude of the magnetic moments modulate) and a helical magnetic phase (where the magnitude is fixed but direction rotates) on a Bravais lattice, polarized neutrons are required. This has been confirmed in Ref. \onlinecite{Stock15:114} and illustrated in Fig. \ref{fig:figure1} $(b)$ with the helical nature evident from a projection along the crystallographic $c$-axis. A transitional phase (identified as ``Phase-II") exists at intermediate fields and temperatures and is defined by an incommensurate wavevector of $\vec{Q}_{0}=({1\over2}, {1\over2}, 0.298)$ with a potentially elliptical structure.~\cite{Raymond07:19}  This intermediate phase will not be discussed in this paper as we do not observe a clear signature of it in the excitations.  A high-field and low temperature phase (``Phase - III") is illustrated in Fig. \ref{fig:figure1} $(c)$ and consists of a commensurate block magnetic order defined by the wavevector $\vec{Q}_{0}=({1\over2}, {1\over2}, 0.25)$.  We will focus our discussion below on the magnetic dynamics in the low temperature Phases-I and III.  The magnetic field thus provides a means to tune the magnetic structure from a helical incommensurate to a commensurate collinear phase. 

In this paper we analyze the magnetic excitations of CeRhIn$_{5}$ in the antiferromagnetic phase which is parent to unconventional superconductivity.  We describe the single magnetic particle excitations using an ``excitonic" model with coupled basis states defined by the single-ion crystal field Hamiltonian which incorporates magnetic anisotropy.  This results in two distinct modes with polarization defined by the crystalline electric field and previously reported with neutron spectroscopy.  We apply a magnetic field to identify strong multiparticle excitations and discuss possible origins of these excitations.

\section{Experiment}

To probe the low-energy magnetic excitations we have applied neutron spectroscopy which is directly sensitive to dipole allowed fluctuations.  In this paper we present data on the magnetic field dependence of the magnetic excitations in CeRhIn$_{5}$.  The sample is the same as used in Ref. \onlinecite{Stock15:114}. The experiments were performed on the MACS triple-axis spectrometer at NIST (Gaithersburg, USA).~\cite{Rodriguez08:19}  Given the need to sample a large region of momentum space, the multi detector arrangement and large flux on the sample was advantageous. Cooled filters of Beryllium were used after the sample with the final energy fixed by a double bounce PG(002) analyzer with E$_{f}$ =5.1 meV and E$_{i}$ defined by a double focused PG(002) monochromator.  The sample was mounted in a 11 T vertical magnetic field such that reflections of the form (HHL) lay within the horizontal scattering plane and cooled to temperatures as low as 0.5 K, well within the antiferromagnetically ordered phase of CeRhIn$_{5}$ ($T_{N}$=3.8 K).   

\section{Results}

\begin{figure}
	\begin{center}
		\includegraphics[width=100mm,trim=3cm 4.0cm 2cm 4.0cm,clip=true]{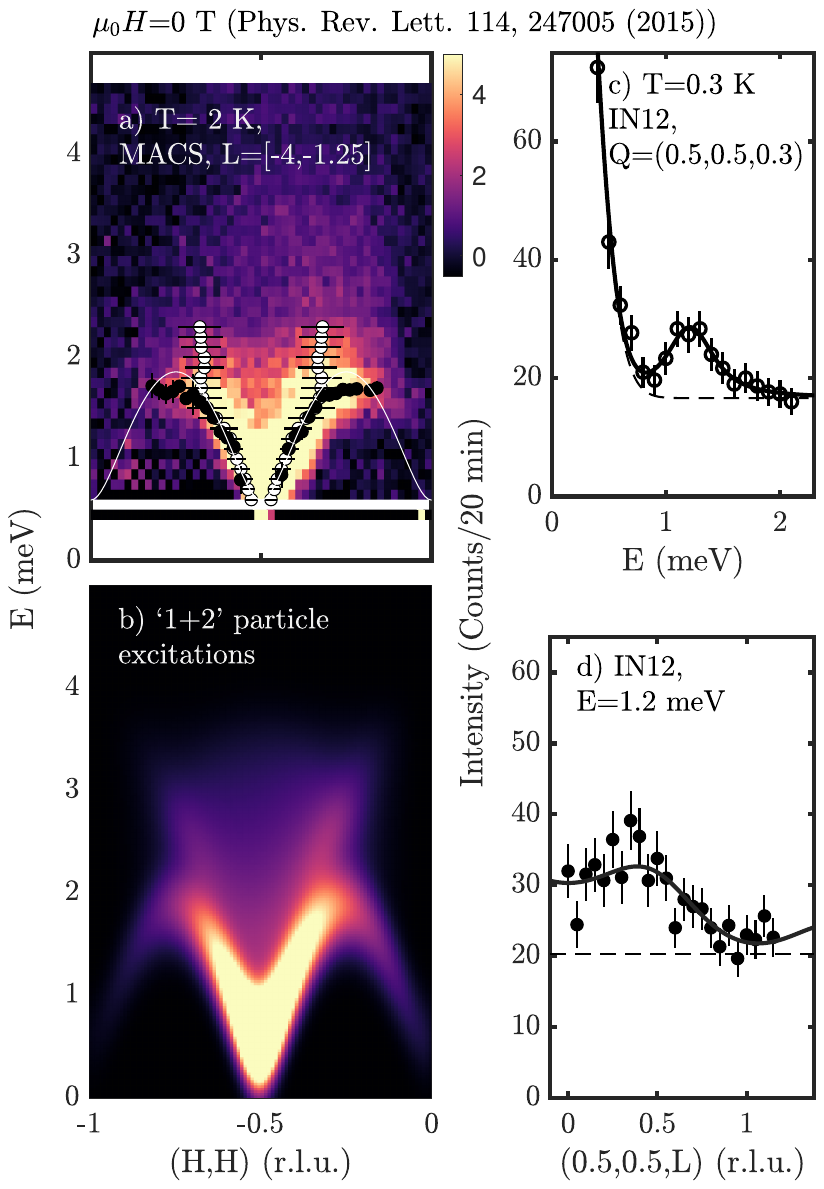}
	\end{center}
	\caption{A summary of the zero field magnetic fluctuations in CeRhIn$_{5}$ below $T_{N}$ previously reported based on neutron spectroscopy. $(a)$ Neutron spectroscopy measurements on MACS compared against a $(b)$ heuristic multimagnon calculation.  A second gapped mode is found in $(c)$ using IN12 which corresponds to fluctuations polarized along the crystallographic $c$ axis as confirmed from a scan of the intensity along [001] shown in $(d)$.}
	\label{fig:zero_field}
\end{figure}

\subsection{Zero-field results summary}

We first review the magnetic excitations at zero applied field and summarized based on our previous results~\cite{Stock15:114} in Fig. \ref{fig:zero_field}.  Fig. \ref{fig:zero_field} $(a)$ illustrates the magnetic excitations measured on the MACS spectrometer (NIST).  These excitations emanate from the commensurate (${1\over 2}$, ${1\over 2}$) position with no observable momentum dispersion along $L$.   The solid black points are the peak positions extracted from one-dimensional constant momentum cuts through the data and the white open circles are from analogous constant energy cuts.  Along with a band of magnetic excitations which is defined by the instrument resolution in momentum and energy (hence well defined and underdamped), a momentum and broadened continuum of excitations is observable up to at least 4 meV.   This is parameterized in terms of multiparticle (or ``2-magnon") in Fig. \ref{fig:zero_field} $(b)$ as discussed in Ref. \onlinecite{Stock15:114}. 

The experimental data in Fig. \ref{fig:zero_field} $(a)$ is integrated over a wide range of momentum transfers along $L$ owing to the two-dimensional nature of the magnetic fluctuations.  Because of selection rules of magnetic neutron scattering  which impose that the cross section is sensitive to magnetic moments $\it{perpendicular}$ to the momentum transfer, Fig. \ref{fig:zero_field} $(a)$ is primarily sensitive to fluctuations within the crystallographic $a-b$ plane.  To probe magnetic fluctuations polarized along the crystallographic $c$-axis, Ref. \onlinecite{Stock15:114} used IN12 (ILL) to study magnetic fluctuations at small $L$ momentum transfers and hence more sensitive to fluctuations polarized perpendicular to the crystallographic $a-b$ plane.  The results are summarized through a constant momentum scan in Fig. \ref{fig:zero_field} $(c)$ and a constant energy scan in Fig. \ref{fig:zero_field} $(d)$ illustrating the decay of intensity with increasing momentum transfer along $L$ indicative that these excitations correspond to fluctuations polarized along the crystallographic $c$-axis.  This is further confirmed by results presented in Ref. \onlinecite{Das14:113} which present time-of-flight data illustrating the presence of at least two distinct underdamped magnetic modes.

The zero field neutron response consists of both single-particle (underdamped and well defined in momentum and energy) and multi-particle (momentum and energy broadened) components.  The single-particle component consists of two bands with a low-energy band dispersing from the commensurate $\vec{Q}=({1\over2},{1\over2})$.  These excitations are polarized within the crystallographic $a-b$ plane.  The higher energy mode illustrated in Fig. \ref{fig:zero_field} $(c)$ corresponds to fluctuations along the crystallographic $c$-axis.
 
\subsection{Results in applied magnetic field}

\begin{figure}
	\begin{center}
		\includegraphics[width=95mm,trim=2.5cm 2.5cm 1cm 3.5cm,clip=true]{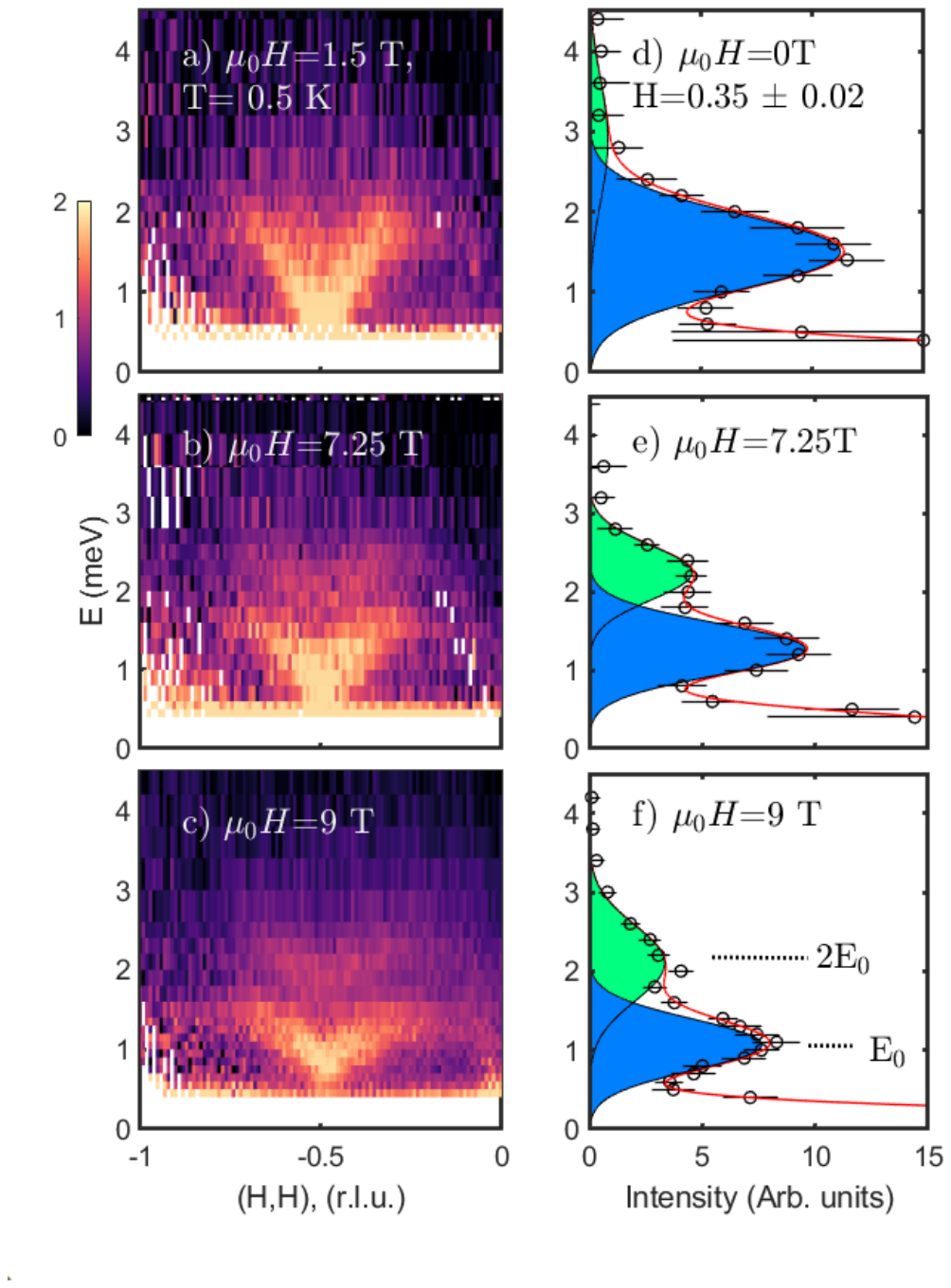}
	\end{center}
	\caption{Evolution of the magnetic field spectrum in a vertical applied magnetic field along the [1$\overline{1}$0] direction.  $(a-c)$ display momentum-energy maps at $\mu_{0}H$=1.5, 7.25, and 9 T.  Panels $(d-f)$ display cuts at each field at H=0.35 $\pm$ 0.02 rlu.}
	\label{fig:applied_field}
\end{figure}

We now discuss the magnetic excitation spectrum at low temperatures in the antiferromagnetic phase with an applied field along the [1$\overline{1}$0] direction with the results summarized in Fig. \ref{fig:applied_field}.  For a small magnetic field of $\mu_{0}H=1.5$ T (Fig. \ref{fig:applied_field} $a$), where the static magnetism remains within the helical ``Phase-I" (Fig. \ref{fig:figure1}), little qualitative change is observed from zero field results (Fig. \ref{fig:zero_field}).  On entering the collinear ``Phase-III" the single particle band becomes clearly gapped at the magnetic zone center, indicative of an anisotropy, as shown in momentum-energy maps at $\mu_{0}H=$ 7.25 and 9 T (Figs. \ref{fig:applied_field} $b,c$).  The bandwidth of the lower energy single band also decreases and this is evidenced on comparison of Fig. \ref{fig:applied_field} $(b)$ ($\mu_{0}H=$7.25 T) and Fig. \ref{fig:applied_field} $(c)$ ($\mu_{0}H=$9 T).  This indicates that the exchange interaction between Ce$^{3+}$ ions is weakened on application of an in-plane magnetic field.  These results are generally consistent with the increase in the anisotropy gap and also the decrease in exchange parameters noted in Ref. \onlinecite{Fobes18:14} for CeRhIn$_{5}$ using chopper spectrometers focusing on in-plane excitations at small values of $L$.

On increasing the magnetic field, the energy and momentum broadened multiparticle excitations observed at zero field in Fig. \ref{fig:figure1} sharpen and become more well defined in momentum and energy.  This is particularly evident in the momentum-energy map in Fig. \ref{fig:applied_field} $(c)$ at $\mu_{0}H=9$ T and the one-dimensional cut at H=0.35 $\pm$ 0.02 r.l.u. in Fig. \ref{fig:applied_field} $(f)$.  There are two modes in the neutron spectra seen at $\mu_{0}H=9$ T with the upper mode being a factor of 2 at higher energies than the lower mode.  This is also evident at $\mu_{0}H=7.25$ T displayed in Fig. \ref{fig:applied_field} $(b)$ and in the one-dimensional cut in Fig. \ref{fig:applied_field} $(e)$.  This second mode can be understood in terms of a multiparticle process based on the lower mode.  With the application of a magnetic field, the energetic gap at the magnetic zone center increases along with the bandwidth of the mode decreasing. This sharpening in energy of the lower single particle branch narrows the allowed multiple particle phase space permitted through exciting two single magnons.  In the limit where there is nearly no momentum dispersion, the single and multiparticle processes would consist of two flat in momentum excitations separated by a factor of 2.  The application of a magnetic field in CeRhIn$_{5}$ clearly shows the existence of a multiparticle process at higher energies with energy and bandwidth tunable using an applied magnetic field.  This demonstrates that the low-energy mode discussed above in the context of the $\mu_{0}H$=0 T data, which is polarized within the crystallographic $a-b$ plane, is unstable to multiparticle processes.  

\section{Theory}

In this section we aim to understand the excitations presented above.  We first discuss various approaches to modeling the magnetic fluctuations and then apply these to the case of CeRhIn$_{5}$ in a magnetic field.  Notably, we compare and contrast the ``effective spin'' and ``excitonic" approaches to this problem that are discussed in the context of the underlying symmetry groups that are important for the magnetic excitations.  SU(2) provides an interesting example of this as the SU(2) group covers O(3) which has a continuous rotational symmetry requiring a gapless Goldstone mode in the magnetic excitation spectrum for a Hamiltonian with SU(2) symmetry.  This is an example where the symmetry group has consequences on the predicted magnetic excitations.  We use Co$^{2+}$ and CoO as an illustrative example to motivate the distinction and need for various approaches for magnetic excitations.~\cite{Sarte19:100,Sarte18:98,Cowley13:88} 

\textit{Effective ``spin" ($j_{eff}$) Spin-Wave Approach}- Magnetic excitations are typically dealt with in the field of scattering in terms of searching out solutions of the Landau-Lifshitz equation for an ``effective'' spin of fixed length under the constraints of a magnetic Hamiltonian which accounts for interactions between spatially separated magnetic ions.  The Hamiltonian is built up from Heisenberg interactions between these effective spins.  We note that the resulting Hamiltonian may not have the same symmetry as these spins with weak anisotropic terms introduced for parameterization purposes.    

This approach can be applied to transition metal compounds containing Co$^{2+}$ with small exchange coupling and hence small molecular field environments.  This has been shown by us in our study of the CoO~\cite{Sarte19:100} where the single-ion Hamiltonian, incorporating crystalline electric field and spin-orbit terms, can be rotated to form a block diagonal matrix.  In terms of group theory, this is identical to formulating the irreducible representations. The ground state Kramers doublet of Co$^{2+}$ in an octahedral crystalline electric field is isomorphic with SU(2) and the underlying Lie algebra generators are proportional to the 2 $\times$ 2 Pauli matrices.~\cite{Sarte18:98}  

The entire single-ion energy spectrum, from a Hamiltonian consisting of crystalline electric and spin-orbit coupling terms in an isolated Co$^{2+}$ ion, results in three energy multiplets with each having a total effective angular momentum (which also defines the degeneracy) of $j_{eff}$=${1 \over 2}$, ${3 \over 2}$, and ${5 \over 2}$.  We note that the constant of proportionality, or the projection factors, that allow mapping onto effective angular momentum operators, are isotropic scalars.  For cases where the ground state $j_{eff}$=${1 \over 2}$ manifold is energetically well separated from higher energy $j_{eff}$ levels an effective Hamiltonian can be constructed for the excitations within only the ground state Kramers doublet based on interacting $S={1 \over 2}$ operators.  This in turn facilitates excitations to be calculated using time evolution defined by the Landau-Lifshitz equation (${d \over {dt}} \langle S \rangle = h_{MF} \times \langle S \rangle$). 

\textit{Interacting multilevel approach}- However, in the case of CoO where, and unusually in transition metal ions, the exchange interaction, spin-orbit coupling, and crystalline electric field distortion parameters are all comparable and the three $j_{eff}$ eigenstates of the single-ion Hamiltonian are strongly mixed.  This invalidates the effective spin $S_{eff}={1\over2}$ approximation discussed above and requires the multi-level single-ion eigenstates to be coupled and accounted for rather than parameterization in the magnetic Hamiltonian.  

We have followed the study of interactions between quasi-particles in electronic models where the single-particle Greens function is fundamental for accounting for interactions.  In the case of insulating magnetic systems, and the approach we take here, this single-particle Green's function is constructed based on the single-ion Hamiltonian to reflect the underlying crystalline electric field.  Such a Hamiltonian naturally accounts for the lattice space group symmetry and indeed the eigenvectors form a representation of the space group.~\cite{Chen:xx}  Instead of the conventional approach utilizing pseudoboson operators in SU(2) as a starting point for perturbation, this approach accounts for the N-multilevel nature and is akin to SU(N) theory developed for magnetic excitations~\cite{Shiina03:72,Hasegawa12:81,Muniz14:2014,Tomiyasu06:75} which brings in other members of the Lie algebra beyond the dipole operators.  

\textit{Case of CeRhIn$_{5}$}- For the rare-earth system of interest here, the magnetic ion is the $4f$ element Ce$^{3+}$. Given that spin-orbit coupling is considered as the dominant energy scale in the magnetic Hamiltonian in $4f$ elements, the observable operators which commutes with the Hamiltonian, and from which the eigenvalues define the good quantum numbers, are the components of the total angular momentum operator $\vec{J}=\vec{L}+\vec{S}$.  In the case of Ce$^{3+}$, the total angular momentum corresponds to $j={5 \over 2}$.  The 6 basis states (that are eigenvectors of the observable $J_{z}$), in the presence of crystalline electric field, break up into doubly degenerate levels.  Since the crystalline electric field is symmetric under time reversal symmetry Kramers' theorem~\cite{Yosida:book} applies and the doublet degeneracy cannot be broken unless a time reversal symmetry breaking field (e.g. Zeeman magnetic or molecular field) is introduced.

\subsection{Effective-spin model}

We first review the conventional effective $S={1\over 2}$ model to parameterize the magnetic excitations.~\cite{Yosida:book}  In this case, we assume that the ground state Kramers doublet is separated in energy from the first excited Kramers doublet by an energy much greater than the temperatures of the experiment.  In such a case, the ground state two-level single-ion doublet can be mapped to a pseudospin $S={1\over2}$ particle in a magnetic field.~\cite{Allen:book}  This approach starts with assuming SU(2) symmetry and parameterizes the Hamiltonian with anisotropy as,

\begin{equation*}
	\mathcal{H}=J\sum_{ij}\left(S_{i}^{x}S_{j}^{x}+S_{i}^{y}S_{j}^{y} +\Delta S_{i}^{z}S_{j}^{z} \right)
\end{equation*}

\noindent where $J$ is the magnetic exchange coupling between neighboring spins defined by the indices $i,j$.  The parameter $\Delta$ defines a comparatively weak anisotropy along the crystallographic $c$ axis.  We note that while the starting operators form an algebra of SU(2), the Hamiltonian does not have SU(2) symmetry and this can be seen by the fact that observable operators $S_{x,y}$ do not commute with the anisotropy term set by the parameter $\Delta$.

The neutron response for magnetic moments of fixed spin $S$ on a two dimensional lattice is well established~\cite{Cowley77:15,Huberman05:72}. The magnon dispersion takes the form,

\begin{equation*}
	E({\bf{q}})=4JS \sqrt{(1+\Delta)^{2}-\gamma^{2}({\bf{q}})}
\end{equation*}

\noindent where $\gamma({\bf{q}})=\cos[\pi(H+K)]\sin[\pi(-K+H)]$.  This parameterization approach to the data in terms of an exchange coupling between nearest neighboring spins was applied in both Ref. \onlinecite{Stock15:114,Das14:113} to describe the low-energy single-particle excitations in CeRhIn$_{5}$.  Ref. \onlinecite{Das14:113} included several different exchange constants to account for both the ordering wavevector and multiple modes.  Ref. \onlinecite{Stock15:114} parameterized the lowest energy excitation, corresponding to in-plane fluctuations above, by a single exchange parameter. Whilst such an approach has real utility for efficiently computing estimates of nearest-neighbor in-plane exchanges, it does not capture the anisotropy of the magnetic excitations without an additional effective parameterization of the magnetic Hamiltonian. 


\subsection{Excitonic models and Kramers doublet magnetism}

Having discussed the effective $S={1\over 2}$ model which is anchored in SU(2) in the previous section, we now apply an excitonic approach including the multi-level nature of the isolated Ce$^{3+}$ ion in a crystalline electric field.  This approach differs from the effective $S={1\over 2}$ model defined above.  Instead of starting by considering isotropic $S={1\over 2}$ SU(2) operators corresponding to a two-level system approximating the ground state Kramers doublet of Ce$^{3+}$, we consider the full $j={5\over 2}$ multiplet with the energetically Kramers degenerate states split by a local crystalline electric field and nearby molecular fields from ordered magnetic moments. These eigenstates of the crystalline electric field Hamiltonian are then coupled using a mean-field random phase approximation (RPA).  We note that this approach involves a starting basis defined by the crystalline electric field which is 6-dimensional (corresponding to $|j\equiv{5\over 2}, m\rangle$ eigenstates of the $J_{z}$ observable in the fundamental representation).  These states are then used as a basis to diagonalize the electrostatic Hamiltonian built from Stevens' operator equivalents.  The resulting single-ion eigenstates are then coupled via dipole operators of angular momentum and we calculate the neutron cross section by applying the Green's function formalism for magnetic dynamics. The approach treats the importance of anisotropy~\cite{Moll17:2} in the heavy fermions on an equal scale with applied field and interactions.

Note that this differs from that of effective $S={1\over 2}$ theories crucially in terms of symmetry properties.  The anisotropy defined by the local crystal electrostatic field is built into the starting multiple single-ion eigenstates.  This is in contrast to parameterizing the Hamiltonian in terms of the anisotropic terms deviating from the symmetric Heisenberg interaction in the effective $S={1\over 2}$ model.   We note that our previous theoretical work on Ca$_{2}$RuO$_{4}$~\cite{Sarte20:102} illustrated the utility of this multi-level approach allowing model terms to be readily mapped to real physical quantities, and also minimizing additional parameters terms in the magnetic Hamiltonian.

\subsubsection{Total magnetic Hamiltonian}

In the following section we pursue this method by first discussing the single-ion Hamiltonian defined by the local crystalline electric field and then the coupling of the eigenstates of thie Hamiltonian using the RPA~\cite{Buyers75:11}.  This procedure requires the total Hamiltonian to be divided into two parts with one describing the single-ion response and the second corresponding to interactions between neighboring magnetic moments.  We define this overall Hamiltonian as

\begin{figure}
	\begin{center}
		\includegraphics[width=85mm,trim=3.5cm 4.3cm 3.7cm 4.3cm,clip=true]{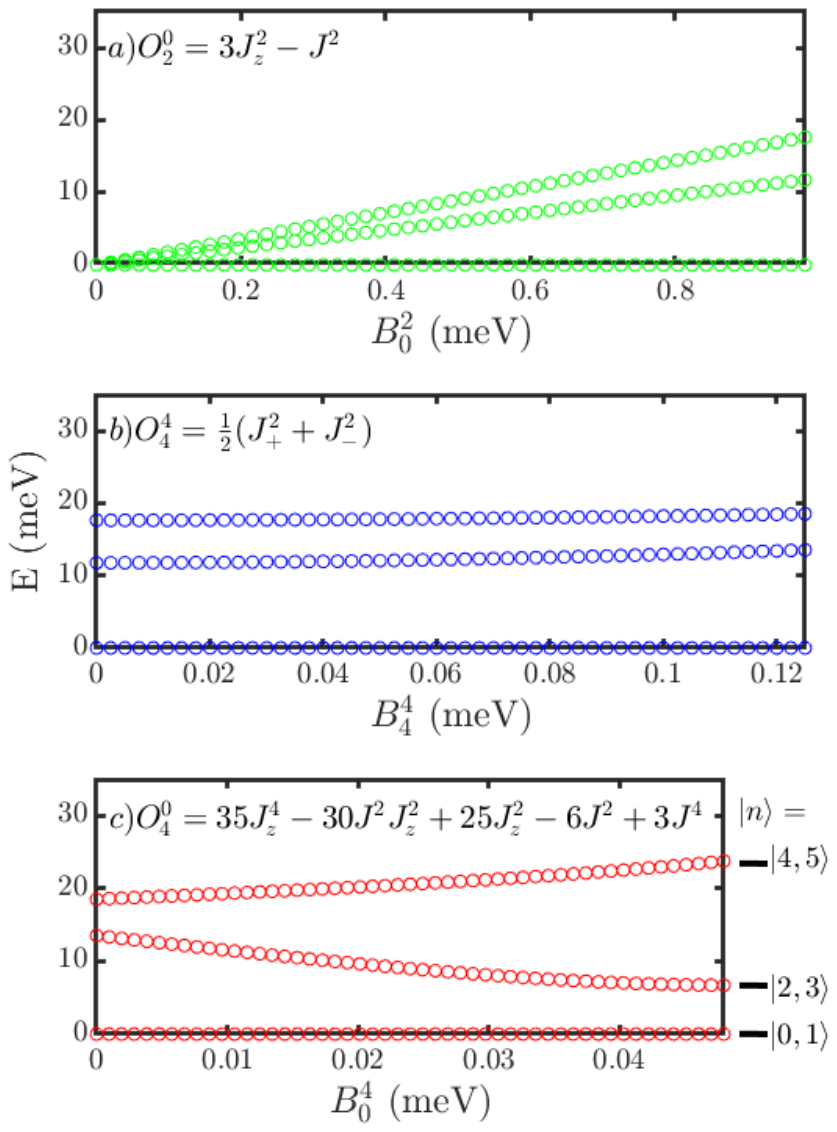}
	\end{center}
	\caption{The energy eigenvalues of the electrostatic contribution to the single-ion Hamiltonian discussed in the text.  In terms of Stevens' operators, the contributions from $(a)$ $O_{2}^{0}$, $(b)$ $O_{4}^{4}$, $(c)$ $O_{4}^{0}$ are illustrated with the final set of 3 Kramers doublets shown.}
	\label{fig:cef_energy}
\end{figure}

\begin{eqnarray}
	\mathcal{H}_{tot}=\mathcal{H}_{SI}+\mathcal{H}_{inter}.
	\label{ham_tot}
\end{eqnarray}

\noindent  We first discuss the single-ion Hamiltonian $\mathcal{H}_{SI}$ and then interactions through $\mathcal{H}_{inter}$ below.  

\subsubsection{Single ion - Local electrostatic and molecular fields}

Based upon symmetry considerations~\cite{Walter84:45} of the point group of the magnetic ion in CeRhIn$_{5}$ (Fig. \ref{fig:figure1}), the electrostatic Hamiltonian for isolated Ce$^{3+}$, $j=5/2$ in a tetragonal crystal field is given by the following equation in terms of Stevens' operators ($\mathcal{O}_{2}^{0}$, $\mathcal{O}_{4}^{0}$, and $\mathcal{O}_{4}^{4}$) and scalar parameters ($B_{2}^{0}$, $B_{4}^{0}$, and $B_{4}^{4}$).

\begin{eqnarray}
	\mathcal{H}_{CEF}=B_{2}^{0}\mathcal{O}_{2}^{0}+B_{4}^{0}\mathcal{O}_{4}^{0}+B_{4}^{4}\mathcal{O}_{4}^{4}
	\label{ham}
\end{eqnarray}

\noindent The Stevens' operators ($\mathcal{O}_{2}^{0}$, $\mathcal{O}_{4}^{0}$, and $\mathcal{O}_{4}^{4}$) are defined~\cite{Hutchings65:45} (reproduced in Table \ref{table_stevens}) in terms of the angular momentum operators $J_{x,y,z}$ ($J_{\pm}\equiv J_{x}\pm i J_{y}$) and $J^{2}$ (which is proportional to the identity) and the variables ($B_{2}^{0}$, $B_{4}^{0}$, and $B_{4}^{4}$) which are real constants that act as parameters for the local crystalline electric field.  We note that the crystal field parameters are the same for the $\mathrm{C}_{4}$, $\mathrm{S}_{4}$, and $\mathrm{C}_{4\nu}$ point group symmetries.~\cite{Walter84:45} For the purposes of the discussion here we will discuss the arguably simplest case of the $\mathrm{C}_{4}$ point group which consists of 4 symmetry operations including the identity operator.  The three Stevens' parameters can be fit from neutron data and also calculated using, for example, point-charge calculations based on Coulomb electrostatic fields.

\begin{table}[ht]
	\caption{Stevens operators relevant for CeRhIn$_{5}$ from Ref. \onlinecite{Hutchings65:45}.}
	\begin{ruledtabular}
		\begin{tabular}{cc}
			$\mathcal{O}_2^0$ & $3J_z^2 - J^2$ \\
			$\mathcal{O}_4^0$ &  $35J_z^4 - 30J^2J_z^2 + 25J_z^2 - 6J^2 + 3(J^2)^2$\\
			$\mathcal{O}_4^4$ & $\frac{1}{2}(J_{+}^4 + J_{-}^4)$ \\
		\end{tabular}
	\end{ruledtabular}
	\label{table_stevens}
\end{table}

The local crystalline electric field in Ce$X$In$_{5}$ has been discussed extensively in previously published papers. We will use the crystal field Stevens' parameters reported in Refs. \onlinecite{Christianson02:66_nov,Willers10:81} for CeRhIn$_{5}$.  We note that both Ref. \onlinecite{Christianson02:66_nov} and Ref. \onlinecite{Willers10:81} give consistent answers for the ground state wavefunctions in CeRhIn$_{5}$ and therefore we have used the average value in our calculations.  These are summarized in Table \ref{table_cef}.  Real space plots of the wavefunctions and a discussion of hybridization is given in Ref. \onlinecite{Sunderman19:99}.

\begin{table}[ht]
	\caption{Stevens coefficients taken from Ref. \onlinecite{Christianson02:66_nov} and \onlinecite{Willers10:81}.  Averages used in our calculations are given in the last column. }
	\centering
	\begin{ruledtabular}
		\begin{tabular}{cccc}
			$B_{l}^{m}$ & Ref. \onlinecite{Christianson02:66_nov} & Ref. \onlinecite{Willers10:81} & Average \\
			\hline
			$B_{2}^{0}$ & -1.03 meV & -0.928 meV & $\equiv$ -0.98 meV \\
			$B_{4}^{0}$ & 0.044 meV & 0.052 meV & $\equiv$ 0.048 meV \\
			$B_{4}^{4}$ & 0.122 meV & 0.128 meV & $\equiv$ 0.125 meV \\
		\end{tabular}
	\end{ruledtabular}
	\label{table_cef}
\end{table}

Diagonalizing the 6 dimensional $\mathcal{H}_{CEF}$ in Eqn. \ref{ham} gives 3 sets of doublets for a total of 6 eigenstates for the crystalline electric field Hamiltonian.  The effect of each of the Stevens' operators in Eqn. \ref{ham} is illustrated in Fig. \ref{fig:cef_energy} with the final energy eigenvalues shown in Fig. \ref{fig:cef_energy} $(c)$. As required by Kramers' theorem, these doublets cannot be split by a time reversal symmetry preserving crystalline electric field, but require a time reversal breaking field to break their doublet degeneracy. Such a field is present in the magnetically ordered phase at low temperatures where neighboring Ce$^{3+}$ magnetic moments impose a molecular field on a given Ce$^{3+}$ site by, 

\begin{eqnarray}
	\mathcal{H}_{MF}= h_{MF} J_{z}
	\label{ham_MF}
\end{eqnarray}

\noindent where the scalar $h_{MF}$ is discussed below when we describe the interaction term of the magnetic Hamiltonian.  We discuss the effect of an applied magnetic field to the sample below in the context of this term in the single-ion Hamiltonian.

The total single-ion Hamiltonian in the low temperature phase is the sum of the crystal field and molecular field terms,

\begin{eqnarray}
	\mathcal{H}_{SI}=\mathcal{H}_{CEF}+\mathcal{H}_{MF}
	\label{ham_SI}
\end{eqnarray}

\noindent which diagonalizing ($\mathcal{H}_{SI}|n\rangle=\omega_{n}|n\rangle$) gives the energy eigenvalue ($\omega_{n}$) and eigenstates ($|n\rangle$) which we couple below.  

\subsubsection{Inter-ion Hamiltonian-Isotropic Heisenberg exchange}

Having presented the single-ion Hamiltonian that defines the multi-level basis states we wish to couple, in this section we discuss the inter-ion coupling of these states.  To describe the inter-ion coupling, we choose to take the simplest approach with the coupling to be bi-linear in the angular momentum operators defined by,

\begin{eqnarray}
	\mathcal{H}_{inter}= {\mathcal{J} \over 2} \sum_{i,j} \vec{J}_{i}\cdot\vec{J}_{j}
	\label{ham_inter}
\end{eqnarray}

\noindent where we take for simplicity $\mathcal{J}$ as a symmetric scalar which is analogous to the Heisenberg exchange discussed above in the context of the $S_{eff}={1\over 2}$ analysis.  We note that this approach naively assumes the exchange coupling between Kramers' doublets on differing sites is the same.  The angular momentum observables are defined by $\vec{J}_{i}$ with $J^{2}=J^{2}_{x}+J^{2}_{y}+J^{2}_{z}$ and $J^{2}|j,m\rangle=j(j+1)|j,m\rangle$ with $j={5\over 2}$ and $J_{z}|j,m\rangle=m|j,m\rangle$.  

While this approach works with a 6 dimensional basis compared to a 2 dimensional $S_{eff}={1\over 2}$ parameterization discussed before, the anisotropy from the local crystalline electric field environment is directly incorporated into this new multi-level basis.  Therefore, anisotropy terms are not used in the interacting Hamiltonian but is rather built into the starting basis derived from the crystal field Hamiltonian discussed above. 

The form we have chosen for the inter-ion Hamiltonian in Eqn. \ref{ham_inter} allows us to write down the following expression for the molecular field $h_{MF}$ in Eqn. \ref{ham_MF},

\begin{eqnarray}
	\mathcal{H}_{MF}= h_{MF} J_{z} = 2 z \mathcal{J}  \langle J_{z} \rangle J_{z} \nonumber
	\label{ham_MF_2}
\end{eqnarray}

\noindent where $z$ is the number of nearest neighbors, $\mathcal{J}$ is the scalar magnetic exchange in Eqn. \ref{ham_inter}, and $\langle J_{z} \rangle$ is the expectation value of the statically ordered magnetic moment on a given Ce$^{3+}$ site.  This time reversal symmetry breaking term in the single-ion Hamiltonian will split the three Kramers doublets illustrated in Fig. \ref{fig:cef_energy}.  We will discuss the effects of this term below in detail in the context of the single-ion susceptibility.

\subsection{Green's functions and relation to neutron scattering}

The energy eigenstates of the single-ion Hamiltonian discussed can be used to directly calculate the neutron scattering response measured in experiment. We summarize the approach we have taken here with a full description given in the appendix.  We note that this methodology has been outlined in several previous papers by us.~\cite{Sarte19:100,Sarte20:102,Lane22:106,Chan23:107,Lane23:5,Riberolles23:14}   We calculate the Green's function ($G^{\alpha \beta} (\bf{q},\omega)$, with $\alpha \beta$ denoting Cartesian coordinates) response which is related to the neutron scattering cross section (by the structure factor $S^{\alpha \beta}({\bf{q}},\omega)$) by the fluctuation-dissipation theorem,  

\begin{equation}
	S^{\alpha \beta}({\bf{q}},\omega)=-\frac{1}{\pi} \frac{1}{1-\exp(\omega/k{\rm{_{B}}}T)} \Im{G^{\alpha \beta} (\bf{q},\omega)}
	\label{Green_to_sqw}
\end{equation}

\noindent The Green's function, in the local rotating frame, $\tilde{G}^{\alpha \beta} (\bf{q},\omega)$, in the case of the 
interaction Hamiltonian stated in Eqn. \ref{ham_inter} is given by the Dyson equation (in the rotating frame denoted via `` $\tilde{}$ " discussed in the Appendix),

\begin{equation}
	\underline{\underline{\tilde{G}}}({\bf{q}},\omega)=\underline{\underline{g}}(\omega)+\underline{\underline{g}}(\omega)\underline{\underline{\tilde{\mathcal{J}}}}({\bf{q}})\underline{\underline{\tilde{G}}}({\bf{q}},\omega)  
	\label{fulleq_Gqw_mat_start}
\end{equation} 

\noindent  The single site susceptibility $g^{\alpha \beta} (\omega)$, where only the ground state $|0\rangle$ is populated and there are no thermally excited states (low temperature $T \rightarrow 0$ limit), is defined as

\begin{equation}
	\begin{split}
		&g^{\alpha\beta}(\omega)=\sum_{n}\frac{\langle 0|J^{\alpha}|n\rangle \langle n|J^{\beta}|0\rangle}{\omega+i\Gamma-(\omega_{n}-\omega_{0})} \\
		&\qquad - \frac{\langle n|J^{\alpha}|0\rangle \langle 0|J^{\beta}|n\rangle}{\omega+i\Gamma+(\omega_{n}-\omega_{0})}
		\label{small_g}
	\end{split}
\end{equation}

\noindent with $|n\rangle$ the eigenstates of the single-ion Hamiltonian with energies $\hbar \omega_{n}$.  

\begin{figure}
	\begin{center}
		\includegraphics[width=85mm,trim=1.25cm 4cm 1.0cm 5cm,clip=true]{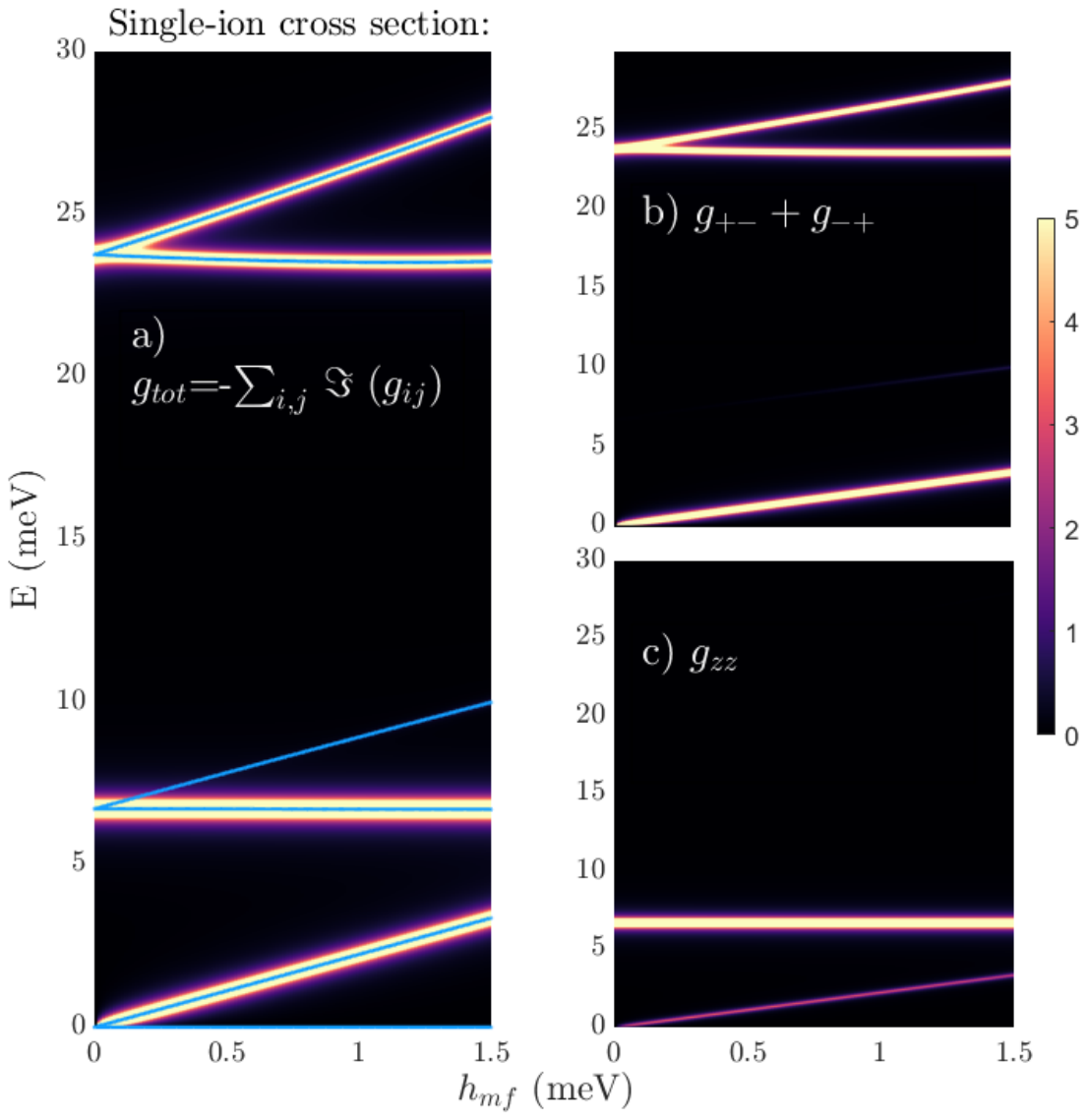}
	\end{center}
	\caption{The single ion susceptibility plotted as a function of molecular field in the single-ion Hamiltonian Eqn. \ref{ham_SI}.  $(a)$ illustrates the sum over all matrix elements $g^{\alpha \beta} (\omega)$ and the anisotropy is illustrated $(b)$ and $(c)$ which plots in-plane and out-of-plane components respectively.  The blue lines in $(a)$ are the energy eigenvalues.  Where $g$ lacks intensity despite an energy eigenvalue implies the excitations to that mode are not allowed by dipole selectrion rules of neutron scattering.}
	\label{fig:small_g}
\end{figure}

The single-site susceptibility provides the foundation for calculating the neutron scattering response with the single-ion states coupled through the exchange Hamiltonian.  The local crystalline electric field imposes an underlying anisotropy to the magnetic response.  This is shown by plotting the total single site response ($=\sum_{\alpha \beta} g^{\alpha \beta}$) as a function of the molecular field term in the Hamiltonian (Eqn. \ref{ham_MF}) which breaks time-reversal symmetry and hence Zeeman splits the three single-ion doublets (\ref{fig:small_g}). 

One advantage of the Green's function approach is that anisotropy is built into the single-ion response ($g^{\alpha \beta}$) originating from the crystalline electric field Hamiltonian.  The full anisotropic single-site susceptibility takes the following matrix form in Cartesian coordinates,

\begin{equation}\label{fourth}
	\begin{split}
		\begin{aligned} \nonumber
			\text{$\underline{\underline{g}}(\omega) = $} \quad &
			\begin{bmatrix} 
				g^{xx} & g^{xy} & g^{xz} \\
				g^{yx} & g^{yy} & g^{yz} \\
				g^{zx} & g^{zy} & g^{zz} \\
			\end{bmatrix}.
		\end{aligned}
	\end{split}
\end{equation}

\noindent  The total single-ion response ($g_{tot}$) is plotted in Fig. \ref{fig:small_g} $(a)$.  The in-plane components ($g^{xx}+g^{yy}$) and out of plane components ($g^{zz}$) are plotted in Fig. \ref{fig:small_g} $(b)$ and $(c)$ as a function of $h_{MF}$ which breaks the doublet degeneracy.  we discuss the value of $h_{MF}$ below. The single-ion response displays anisotropy with differing excitations (particularly at $\sim$ 5 meV) for in-plane ($g^{xx}+g^{yy}$) and out-of-plane ($g^{zz}$) components.  Owing to the anisotropy from the crystalline electric field, the single-ion response is very anisotropic and we will show below that this translates into a spatially anisotropic multi-mode response in the neutron cross section.

With the single site susceptibility discussed, the total neutron cross section can be calculated from the Green's function response by solving the Dyson equation Eqn. \ref{fulleq_Gqw_mat_start}.

\begin{equation}
	\underline{\underline{\tilde{G}}}({\bf{q}},\omega)=\underline{\underline{g}}(\omega) \left[ \underline{\underline{\mathds{1}}}-\underline{\underline{\tilde{\mathcal{J}}}}({\bf{q}}) \underline{\underline{g}}(\omega) \right]^{-1}.\nonumber
\end{equation}

\noindent Through taking the imaginary part of this matrix equation and transforming back to the laboratory frame from the local rotating frame (see Appendix), the total neutron cross section can be calculated.  

\subsection{Comparison with SU(N) spin-wave theory}

The application of Green's functions to neutron scattering involves two key steps.  First, the calculation of the single-site eigenstates based on a local crystalline electric and molecular field Hamiltonian.  The symmetry of these states satisfies a discrete $\mathrm{C}_{4}$ symmetry imposed by the local tetragonal point group symmetry as the three Stevens' operators that define the local crystalline electric field are invariant under the four symmetry operators (which includes the identity) of the $\mathrm{C}_{4}$ group.  The second step of our excitonic model is to couple these single-ion eigenstates through the Random Phase Approximation (RPA). 

We compare this to the general SU(N) spin-wave theory~\cite{Muniz14:2014} which has been previously applied in several compounds including spin-dimer compounds~\cite{Hasegawa12:81} and also in materials where spin-orbit coupling is relevant~\cite{Dong18:97,Joshi99:60,Li98:81}.  Theoretically, our excitonic approach and SU(N) give the same results.  However, there are several differences in the starting point of these theoretical analyses which we outline. In the case of Ce$^{3+}$, there are 6 basis states corresponding to eigenstates of the observable operator $J_{z}$.    The rotation matrix that rotates from the $N$ many $|j,m\rangle$ states to those of the crystal field Hamiltonian belongs to the SU(N$\equiv$6) group.   Semiclassical coherent states~\cite{Perelmov71:26,Gilmore72:74,Rowe93:2,Rowe:book2004,Zhang21:104}  which locally~\cite{Dahlbom23:xx} have continuous SU(N) symmetry are then constructed to describe the members of the 6 dimensional basis.  We note that SU(N$\equiv$6) requires the identity (Casimir) and N$^{2}$-1 (35) generators of symmetry which form the continuous Lie group and hence can be used to describe multipolar excitations.  An arbitrary local order parameter is always contained in the most general SU(N) order parameter with N$^{2}$-1 components.  Recent examples of applications of this can be found in Ref. \onlinecite{Dahlbom24:109}.

SU(N) and our Green's function method aim to move beyond standard spin-wave theory which is based on a starting symmetry of SU(2), like outlined above for the effective $S={1\over2}$ description of CeRhIn$_{5}$. While the interaction Hamiltonian $\mathcal{H}_{inter}$ has SU(2) symmetry with the generators of the Lie Algebra being $J_{x,y,z}$, our excitonic approach applying Green's functions forces the local discrete $\mathrm{C}_{4}$ symmetry as it applies eigenstates of the Stevens' operators that are invariant under the local point group.  We will show below this naturally builds in anisotropy into the predicted excitations.  As illustrated in our previous works~\cite{Sarte20:102} it also allows a direct connection between measurable parameters and physical quantities (such as spin-orbit coupling in the context of transition metal ions).
 
In the approach below we apply the excitonic Green's response function methodology to this problem and do not use the generalized pseudo boson coherent state methodology of SU(N).  In our case the single-ion eigenstates of the local discrete point group are used as a starting basis and then subsequently coupled through the RPA outlined above and provided more in detail in the Appendix.

\begin{figure}
	\begin{center}
		\includegraphics[width=90mm,trim=2.8cm 5cm 2cm 5cm,clip=true]{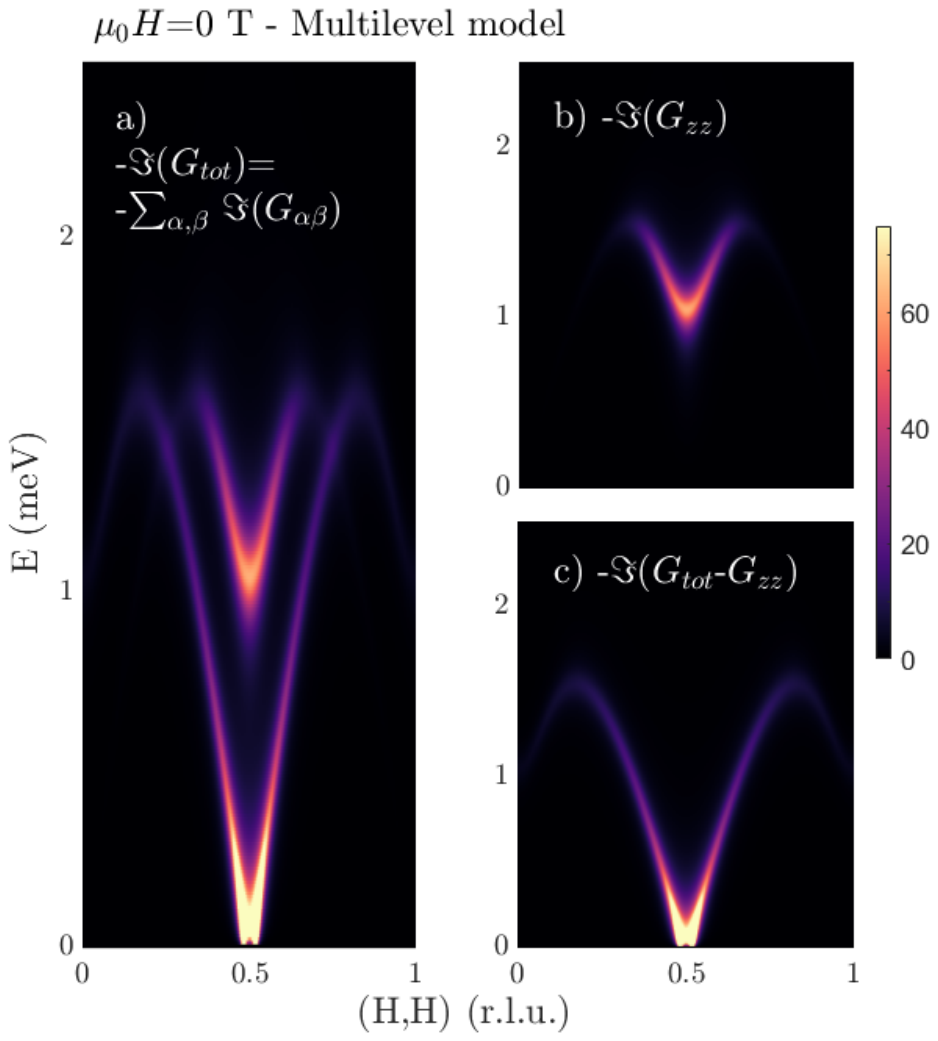}
	\end{center}
	\caption{The calculated $\mu_{0}H=0$ T neutron response.  $(a)$ displays the total cross section and $(b)$ and $(c)$ display the in-plane out and out of plane components. The parameters used for this calculation include the crystalline electric field parameters listed in Table \ref{table_cef}, $\mathcal{J}$=0.135 meV, and $h_{MF}$=0.63 meV (discussed in the text).}
	\label{fig:cross_section_0T}
\end{figure}

\section{Theoretical neutron results}

We calculate the full neutron response for CeRhIn$_{5}$ in Fig. \ref{fig:cross_section_0T} at $\mu_{0}H=0$ T.  We use the values listed in Table. \ref{table_cef} for the Stevens' parameters to describe the crystalline electric field.  We have tweaked $\mathcal{J}$=0.135 meV to match the dispersion bandwidth observed in experiment.  We note that previous $S_{eff}={1\over2}$ models predict a much larger exchange constant of $\sim$0.7-0.9 meV.~\cite{Das14:113,Stock15:114}  However, given that we are considering a full $j={5\over 2}$ model the value for the exchange quoted here should be scaled by a factor of 5, making it in agreement with previous estimates.  For the molecular field component of the single-ion Hamiltonian, given that the $z$-axis is defined through the electrostatic field and the magnetic moments order within $a-b$ plane, we have taken the case of the molecular field to be along $x$.  The molecular field $h_{MF}$ depends on the expectation value $\langle J_{\tilde{z}} \rangle$ which is related to the ordered magnetic moment found in neutron diffraction via $\mu=g_{J}\langle J_{\tilde{z}} \rangle \mu_{B}$, with $g_{J}={6 \over 7}$ for Ce$^{3+}$.  The ordered magnetic moment for CeRhIn$_{5}$ has been reported to be significantly reduced over expectations based on an isolated Ce$^{3+}$ with values of $\sim$ 0.4 $\mu_{B}$ being found.~\cite{Bao00:62,Christianson02:66_aug}  We note that in other Ce$^{3+}$-based compounds such as CeIn$_{3}$~\cite{Benoit80:34,Lawrence80:22} and Ce$_{2}$RhIn$_{8}$~\cite{Bao01:64} the ordered magnetic moment is reported to be $\mu\sim 0.5\mu_{B}$. We can also estimate the ordered moment based on the crystalline electric field Stevens' parameters listed in Table \ref{table_cef}

\begin{equation}
	g_{J}|\langle 0| J_{\tilde{z}} | 0 \rangle| \mu_{B}= 0.49 \mu_{B}. \nonumber
\end{equation}

\noindent This is in good agreement with the values reported for Ce$^{3+}$ heavy fermion compounds. For the purposes of our calculation we have taken an ordered moment of $\mu=0.5 \mu_{B}$ and therefore $h_{MF}={\mu \over {g_{J}\mu_{B}}}\times2\times z \times \mathcal{J}$=0.63 meV (with $z=4$ for the number of nearest neighbors in the $a-b$ plane).  

The calculation predicts two excitations (Fig. \ref{fig:cross_section_0T}) with one being energetically gapless and corresponding to fluctuations within the $a-b$ plane and a second energetically gapped for excitations out of the plane or along the crystallographic $c$-axis.  The gapless excitation is required owing to the presence of a continuous symmetry related to rotating the Ce$^{3+}$ moments within the $a-b$ plane.  This is in qualitative agreement with neutron spectroscopy results for the energetically well defined portion of the neutron response.~\cite{Das14:113,Stock15:114}  The use of excitations from single-ion states incorporates the anisotropy of the lattice and this is reflected in the neutron scattering response and our Green's function calculations.  We note the presence of two distinct modes is consistent with a symmetry analysis of the magnetic ground state which, based on the space group $P4/m$ $2/m$ $2/m$ (No. 123) results in two irreducible representations.  One has two basis vectors within the crystallographic $a-b$ plane and a second irreducible representation which as a single vector along $c$.~\cite{Stock15:114}

\begin{figure}
	\begin{center}
		\includegraphics[width=85mm,trim=2.75cm 4.5cm 2.5cm 5.3cm,clip=true]{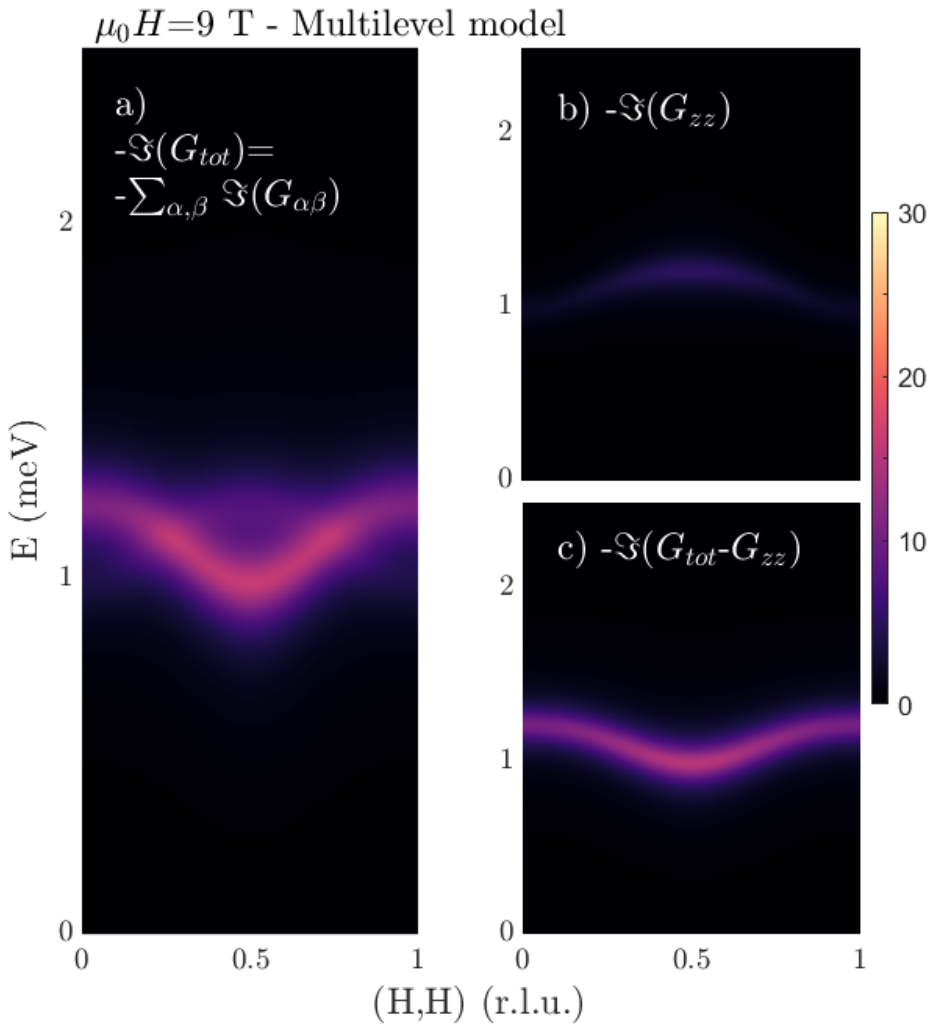}
	\end{center}
	\caption{The calculated $\mu_{0}H=9$ T neutron response.  $(a)$ displays the total cross section and $(b)$ and $(c)$ display the in-plane out and out of plane components. The parameters used for this calculation include the crystalline electric field parameters listed in Table \ref{table_cef}, $\mathcal{J}$=0.03 meV, and $h_{MF}$=0.14 meV (discussed in the text).}
	\label{fig:cross_section_9T}
\end{figure}

The calculated neutron response in an applied field within the crystallographic $a-b$ plane is shown in Fig. \ref{fig:cross_section_9T}.  We have tuned $\mathcal{J}$=0.03 meV so that dispersing band width of the excitation matches the data taken at $\mu_{0}$H= 9 T from MACS in Fig. \ref{fig:applied_field} $(c)$ and $(f)$.  We have then fixed the molecular field based on $\mathcal{J}$ to be $h_{MF}$=0.14 meV.  The applied field along the [1$\overline{1}$0] is included as an additional Zeeman term analogous to the molecular field,

\begin{equation}
	\mathcal{H}_{Zeeman}=\alpha g_{J}\mu_{B}H{(J_{x}+J_{y}) \over {\sqrt{2}}}. 
	\nonumber
\end{equation}

\noindent  We have included a heuristic screening term of $\alpha$=0.88 to account for screening of the magnetic field and field inhomogeneities or misalignment of the magnetic field.  This term was tuned so that the in-plane excitation appears at $\sim$ 1 meV.   The model reproduces the decrease in bandwidith and the enhanced gap with an applied magnetic field and indicates that the apparent increase in anisotropy with applied field is Zeeman driven. The connection between this increase in anisotropy and and large field ($\sim$ 28 T) field-induced density waves~\cite{Ronning17:548} is not clear from this current comparatively low-field study, however the anisotropy gap in the 115 heavy fermion system is important for tuning the ground state~\cite{Willers15:112}.   Interestingly, the model here also predicts a weakening of the the mode polarized along the crystallographic $c$ axis.  This could provide the origin of the fact that this mode was not observable on MACS at small values of momentum transfers along the $c$ axis in our experiments.  

\section{Discussion and conclusions}

In summary, we have applied a general multi-level spin wave model using Green's functions to calculate the neutron scattering response.  This approach brings in  anisotropic crystalline electric field Stevens' operators based on the discrete $\mathrm{C}_{4}$ local point group and the coupling terms via dipole operators in the magnetic Hamiltonian.  This mean-field theory describes the two excitations observed with neutron scattering and also their differing polarization at zero applied field and also can be used to heuristically understand the applied magnetic field dependence.  

In terms of zero field results, the low-energy mode is the gapless Goldstone mode required from the presence of a continuous symmetry of the helical magnetic structure.  Indeed, the entire magnetic structure is symmetric for two-dimensional rotations in the $a-b$ plane.  This corresponds to O(2) operators commuting with the magnetic Hamiltonian.  However, experimentally there has been the suggestion of a gap in this response.  While higher resolution data maybe required to fully confirm this, it is possible that other processes like order-disorder, defects, or crystal field fluctuations maybe the origin.~\cite{Rau16:93}  These are not accounted for in our mean field excitonic model.

It is interesting to relate these excitations to those observed in isostructural superconducting counterparts like CeCoIn$_{5}$.  To our knowledge, in superconducting compounds there has not been a report of temporally well defined (underdamped) spin excitations $\textit{polarized}$ within the crystallographic $a-b$ plane either in superconducting or non-superconducting normal phases.  However, excitations polarized along the crystallographic $c$-axis have been reported in Hg- doped CeCoIn$_{5}$~\cite{Stock18:121} being a precursor to magnetic order~\cite{Bao09:79}.  Similar static results have been reported for Cd doped.~\cite{Nicklas07:76,Gofryk12:109}  No in-plane excitations were observable in the neutron experiments.  Notably, the spin resonance found in superconducting and paramagnetic CeCoIn$_{5}$ was found to also be aniostropic and polarized along the crystallographic $c$-axis.  Comparing antiferromagnetic CeRhIn$_{5}$ with isostructural superconducting CeCoIn$_{5}$ seems to indicate robustness of magnetic excitations polarized along the crystallographic $c$-axis in comparison to the in-plane $a-b$ plane fluctuations.  

The resonance mode has been found to disperse in momentum leading to the suggestion that it may not originate from a $d$-wave superconducting order parameter but rather to the removal of decay channels that cause dampening and hence broadening in energy.~\cite{Song16:7,Song20:3}  This ``magnon" scenario has also been suggested theoretically.~\cite{Chubukov08:101}  Also noteworthy is that the single-ion framework would predict magnetic excitations in the absence of a time reversal symmetry breaking field to be doublets owing to Kramers' theorem.  This is also reported in experiment for the spin resonance where the splitting of this doublet resonance with a field can be used to condense a static spin-density wave ``Q-phase".~\cite{Kenzelmann08:321,Kenzelmann10:104,Blackburn10:105,Koutroulakis10:104} However, we note that idea of attributing the resonance to a magnon has a problem in our current framework described above to model the excitations in CeRhIn$_{5}$.  The gapping of the mode originates from a time reversal symmetry breaking field (here the molecular field).  This is not the case for the CeCoIn$_{5}$ resonance as such a situation would not result in a doublet which seems to imply via Kramers theorem that time reversal symmetry is preserved.  The $d$-wave superconducting order parameter or the magnon description of the resonance still needs further theoretical work.

The instability of the in-plane magnetic excitations is also seen in parent CeRhIn$_{5}$ (illustrated in Fig. \ref{fig:applied_field}) where these excitations are found to coexist with a momentum and energy broadened component that is reminiscent of multi magnon processes reported in insulating model magnets.  This indicates an apparent instability of in-plane excitations in comparison to out-of-plane components.  The origin of this instability and also the apparent stability of $c$-axis magnetic excitations will require further theoretical work along with their link to superconductivity as displayed by the observation of a $c$-axis polarized spin resonance in CeCoIn$_{5}$.

In conclusion, we have analyzed the magnetic excitations in CeRhIn$_{5}$ in terms of a mean field excitonic model constrained by the local $\mathrm{C}_{4}$ point group symmetry.  This captures the anisotropy resulting in two magnetic modes with differing polarization, in agreement with experiment.  We then discuss this framework in the context of an applied magnetic field which controls the local single-site anisotropy.  

\section{Acknowledgements}

This work was funded by the Carnegie Trust for the Universities of Scotland, the Royal Society, the Royal Society of Edinburgh, the STFC, and the EPSRC. Part of this work was carried out at the Brookhaven National Laboratory which is operated for the U. S. Department of Energy by Brookhaven Science Associates (DE-AcO2-98CH10886). H.L acknowledges funding from the Royal Commission for the Exhibition of 1851. 

\section{Appendix: Green's functions and neutron scattering}

In this appendix, we outlined the Green's function approach in detail. 

\subsubsection{Green's functions}

Having defined the basis states built from single-ion physics and the Hamiltonian for interactions in the main paper, we discuss in this Appendix the calculation of neutron scattering intensities and the coupling of these single-ion states at a mean-field level.  To accomplish this we apply the Green's function~\cite{Zubarev60:3} formalism for neutron scattering.

The technique for applying Green's functions to neutron spectroscopy data has been outlined in several papers by us in the context of spin-orbit coupling and $3d$ transition metal ions.  We take the same approach here, but with a differently defined single-ion Hamiltonian which determines the eigenstates that are coupled at a mean field level.  Our initial studies investigated collinear magnetically ordered compounds, such as spin-orbit coupling in CoO.  More recently we have generalized this approach to non-collinear compounds and reduced the number of equations that are required for the calculations. We have further demonstrated the equivalence with conventional spin-wave theory, such as can be applied in \textit{SpinW}~\cite{Toth15:27}, through studying compounds which lack an orbital degree of freedom and have spin-only magnetic moments.  In the formalism that follows, we assume one magnetic ion per unit cell, taking advantage of the rotating frame, and that we are at low temperatures where the excited levels are not thermally populated.

We will merely summarize the key calculation results here, but for more details see Ref. \onlinecite{Lane21:104}. In this analysis we make use of summation notation requiring multiple sets of indices corresponding to different entities and also different labelling of operators.  In Table \ref{Table:indices} we define our notation.

\begin{table}[h]
	\caption{\label{Table:indices} Summary of labeling convention for indices and notation.}
	\begin{ruledtabular} \label{Table_notation}
		\begin{tabular}{cc}
			Index & Description \\ 
			\hline 
			$i$, $j$, .. & unit cell position indices  \\ 
			$\alpha$, $\beta$, $\mu$, $\nu$,... & Cartesian coordinates $(x,y,z)$  \\ 
			$\underline{\underline{X}}$ & Matrix quantities \\
			$A, B, C,...$ & Quantities in lab frame \\
			$\tilde{A}, \tilde{B}, \tilde{C},...$ & Quantities in local rotating frame \\ 
			$J_{i}$ & Angular momentum observables \\
			$\mathcal{J}$ & Symmetric (scalar) Heisenberg exchange \\ 
		\end{tabular}
	\end{ruledtabular}
\end{table} 

The neutron scattering cross section is directly proportional to the structure factor $S({\bf{q}},\omega)$

\begin{equation*}
	S({\bf{q}},\omega)=g_{L}^{2}f^2({\bf{q}})\sum_{\alpha \beta} (\delta_{\alpha \beta}-\hat{q}_{\alpha}\hat{q}_{\beta}) S^{\alpha \beta}({\bf{q}},\omega), 
\end{equation*}

\noindent where $g_{L}$ is the Landé factor and $f({\bf{q}})$ is the Ce$^{3+}$ magnetic form factor. The summation is over $\alpha, \beta = x,y,z$ and 

\begin{equation*}
	S^{\alpha\beta}({\bf{q}},\omega)=\frac{1}{2\pi} \int dt e^{i\omega t} \langle J^{\alpha} ({\bf{q}},t) J^{\beta}(-{\bf{q}},0) \rangle.
\end{equation*}

\noindent The relation to the Green's response function which can be directly connected with the neutron scattering intensity via the fluctuation-dissipation theorem is given by,

\begin{equation}
	S^{\alpha \beta}({\bf{q}},\omega)=-\frac{1}{\pi} \frac{1}{1-\exp(\omega/k{\rm{_{B}}}T)} \Im{G^{\alpha \beta} (\bf{q},\omega)}
	\label{Green_to_sqw}
\end{equation}

\noindent with $T$ the temperature $k_{B}$ the Boltzmann factor.  The Green's function here is defined in terms of the commutator of the angular momentum observables 

\begin{equation}
	G^{\alpha\beta}_{i,j}(t)=-i\Theta(t)\langle [J^{\alpha}_{i}(t),J^{\beta}_{j}(0)]\rangle
	\nonumber
\end{equation}

\noindent with the Heaviside function $\Theta(t)$ forcing causality.  Through taking a time derivative of the Green's function, an equation of motion can be derived which depends on the commutator of the angular momentum operators with the magnetic Hamiltonian discussed above.  These commutators are discussed in previous references and after applying a mean field decoupling scheme~\cite{Buyers75:11}, the following expression for the Green's function can be derived.

\begin{equation}
	\begin{split}
		&G^{\alpha\beta}_{i,j}(\omega)=g^{\alpha\beta}(\omega)\delta_{ij}\\ &\qquad+\sum_{\nu}\sum_{k}g^{\alpha\nu}(\omega)\mathcal{J}_{ik}G_{kj}^{\nu\beta}(\omega).
	\end{split}
	\label{fulleq_G}
\end{equation} 

\noindent  The single site susceptibility $g^{\alpha \beta} (\omega)$, where only the ground state is populated and there are no thermally excited states (low temperature $T \rightarrow 0$ limit), is defined as

\begin{equation}
	\begin{split}
		&g^{\alpha\beta}(\omega)=\sum_{n}\frac{\langle 0|J^{\alpha}|n\rangle \langle n|J^{\beta}|0\rangle}{\omega+i\Gamma-(\omega_{n}-\omega_{0})} \\
		&\qquad - \frac{\langle n|J^{\alpha}|0\rangle \langle 0|J^{\beta}|n\rangle}{\omega+i\Gamma+(\omega_{n}-\omega_{0})}
		\label{small_g}
	\end{split}
\end{equation}

\noindent where $\Gamma$ is required to ensure analyticity and can be physically interpreted as an energy linewidth from the spectrometer resolution.  The summation over the states $|n\rangle$ which are the coherent states defined by the single-ion Hamiltonian discussed above.  

We note that Eqn. \ref{fulleq_G} has the form of the Dyson equation with the single site susceptibility having the role of the bare propagator and the second term a first order perturbative correction.~\cite{Lane21:104}  Indeed, the Dyson equation (Eqn. \ref{fulleq_G}) can then be written more compactly in matrix form which allows Cartesian coordinates to be included implicitly while summations and indices are only over unit cells.  Using matrix notation, the Dyson equation in terms of atomic coordinates ${ij}$ and frequency $\omega$ takes the form,

\begin{equation}
	\underline{\underline{G}}_{i,j}(\omega)=\underline{\underline{g}}(\omega)\delta_{ij}+\sum_{k}\underline{\underline{g}}(\omega)\underline{\underline{\mathcal{J}}}_{ik}\underline{\underline{G}}_{kj}(\omega) \nonumber.
\end{equation} 

\noindent The Heisenberg coupling here, while written as a matrix, will be approximated as proportional to the 3$\times$3 identity, $\doubleunderline{\mathcal{J}}_{ik}\equiv \mathcal{J}_{ik} \doubleunderline{\mathds{1}}$, where $\mathcal{J}_{ik}$ is a scalar. Our Dyson-equation is an equation of motion for a quasi-particle describing excitations between single-ion states in a crystalline electric field.

\subsubsection{Rotating frame for incommensurate magnetic structures}

The above description for the Green's functions was cast in terms a single magnetic site per unit cell.  However, in the case of CeRhIn$_{5}$ the magnetic structure is incommensurate resulting from a helical (within the $a-b$ plane) magnetic structure with a periodicity along the $c-$axis defined by the propagation wavevector $\vec{Q}_{0}=({1\over 2},{1 \over 2},0.297)$.  Given the extended nature of the magnetic unit cell, it is not possible to define a crystallographic basis that can be translated and the incommensurate nature of the magnetic structure results in the magnetic ion in each unit cell having a unique single site susceptibility ($\underline{\underline{g}} (\omega)$).  To circumvent this problem and avoid using large unit cells which are cumbersome for calculations, we therefore utilize a rotating frame formalism which allows us to work in the primitive unit cell and also considerably reduces the number of equations required to be solved numerically.

In this approach, we move to a reference frame that rotates with the magnetic structure, $(\tilde{x},\tilde{y},\tilde{z})$ with the magnetic moment oriented along $\tilde{z}$ in its local reference frame.  This process has been outlined in general in Ref. \onlinecite{Lane22:106} for the case of RbFe$^{2+}$Fe$^{3+}$F$_{6}$.  To apply this, we need to define a rotation matrix that rotates this local coordinate system from one Ce$^{3+}$ site to another.  This is most conveniently done utilizing Rodrigues' rotation formula which allows rotation matrices to be constructed from the propagation wavevector (in this case ${\bf{Q}}=(0.5, 0.5, 0.297)$).  The rotation matrix for a site at a position ${\bf{r_{i}}}$ can be written 

\begin{subequations}
	\begin{align}
		\label{Rodrigues}
		\underline{\underline{R}}_{i}=e^{i\mathbf{Q}\cdot \mathbf{r}_{i}}\underline{\underline{\Phi}} +e^{-i\mathbf{Q}\cdot \mathbf{r}_{i}}\underline{\underline{\Phi}}^{*}+\mathbf{n}\mathbf{n}^{T}\\
		\underline{\underline{\Phi}}=\frac{1}{2}\left(\underline{\underline{\mathds{1}}}-\mathbf{n}\mathbf{n}^{T}-i\underline{\underline{[n]}}_{\times}\right). \label{Rodriguesb}
	\end{align} 
\end{subequations}

\noindent where ${\bf{n}}$ is a vector normal to the plane of the spiral (in the case of CeRhIn$_{5}$ this is ${\bf{n}}=(0,0,1)^{T}$).  In matrix form

\begin{equation*}
	\begin{split}
		\begin{aligned}
			\text{$\underline{\underline{[n]}}_\times =$ } \quad &
			\begin{bmatrix}
				0 & -n_k & n_j \\
				n_k & 0 & -n_i \\
				-n_j & n_i & 0 \\
			\end{bmatrix}.
		\end{aligned}
	\end{split}
\end{equation*}

\noindent Using the Eqn. \ref{Rodrigues} and \ref{Rodriguesb} to define a rotation matrix for each site, we can define a rotation matrix for sites ${ij}$ and write the Dyson equation for the Green's function $\tilde{G}$ in the rotating frame

\begin{equation}
	\begin{split}
		&\underline{\underline{\tilde{G}}}_{i,j}(\omega)=\underline{\underline{R}}_{i}^{T}\underline{\underline{G}}_{i,j}(\omega)\underline{\underline{R}}_{j}=\underline{\underline{g}}(\omega)\delta_{ij}...\\ &\qquad+\sum_{k}\underline{\underline{g}}(\omega)\left(\underline{\underline{R}}_{i}^{T}\underline{\underline{\mathcal{J}}}_{ik}\underline{\underline{R}}_{k}\right)\left(\underline{\underline{R}}_{k}^{T}\underline{\underline{G}}_{kj}(\omega)\underline{\underline{R}}_{j}\right) \nonumber \\
	\end{split}
	\nonumber
\end{equation} 

\noindent where we have used the unitary property of rotation matrices $\underline{\underline{R}}_{k}\underline{\underline{R}}_{k}^{T}\equiv\underline{\underline{\mathds{1}}}$.  We have also used the approximation that the rotation matrices ($\underline{\underline{R}}_{i}$) commute with the single-ion susceptibility matrices $\underline{\underline{g}}(\omega)$.~\cite{Jensen:book}  This approximation will be discussed and quantified below in the context of the specific case of CeRhIn$_{5}$.  Putting this all together and collecting the terms in brackets, we can write the Dyson equation in the rotating frame as

\begin{equation}
	\begin{split}
		&\underline{\underline{\tilde{G}}}_{i,j}(\omega)=\underline{\underline{g}}(\omega)\delta_{ij}...\\ &\qquad+\sum_{k}\underline{\underline{g}}(\omega)\underline{\underline{\tilde{\mathcal{J}}}}_{ik}\underline{\underline{\tilde{G}}}_{kj}(\omega). \nonumber
	\end{split}
	\label{fulleq_G_matrix}
\end{equation} 

\noindent  Note we have used the notation that all quantities written in the rotation frame have a ``tilde" (for example $\underline{\underline{\tilde{G}}}$ or $\underline{\underline{\tilde{J}}}$)
above them.  Taking the Fourier transform and switching from index notation to matrix notation (for example $G^{\alpha\beta} \rightarrow \underline{\underline{G}}$) gives

\begin{equation}
	\underline{\underline{\tilde{G}}}({\bf{q}},\omega)=\underline{\underline{g}}(\omega)+\underline{\underline{g}}(\omega)\underline{\underline{\tilde{\mathcal{J}}}}({\bf{q}})\underline{\underline{\tilde{G}}}({\bf{q}},\omega)  
	\label{fulleq_Gqw_mat}
\end{equation} 

\noindent  Regarding the Fourier transform of the exchange term $\underline{\underline{\mathcal{J}}}$ which couples differing magnetic Ce$^{3+}$ sites,  we define $\underline{\underline{\mathcal{J}}}$ as

\begin{equation}
	\underline{\underline{\mathcal{J}}}(\mathbf{q})=\sum_{mn}\underline{\underline{\mathcal{J}}}_{mn}e^{-i\mathbf{q}\cdot(\mathbf{r}_{m}-\mathbf{r}_{n})}\label{Jmat:eq}\\
\end{equation}

\begin{figure}
	\begin{center}
		\includegraphics[width=90mm,trim=3cm 4.5cm 3cm 5.5cm,clip=true]{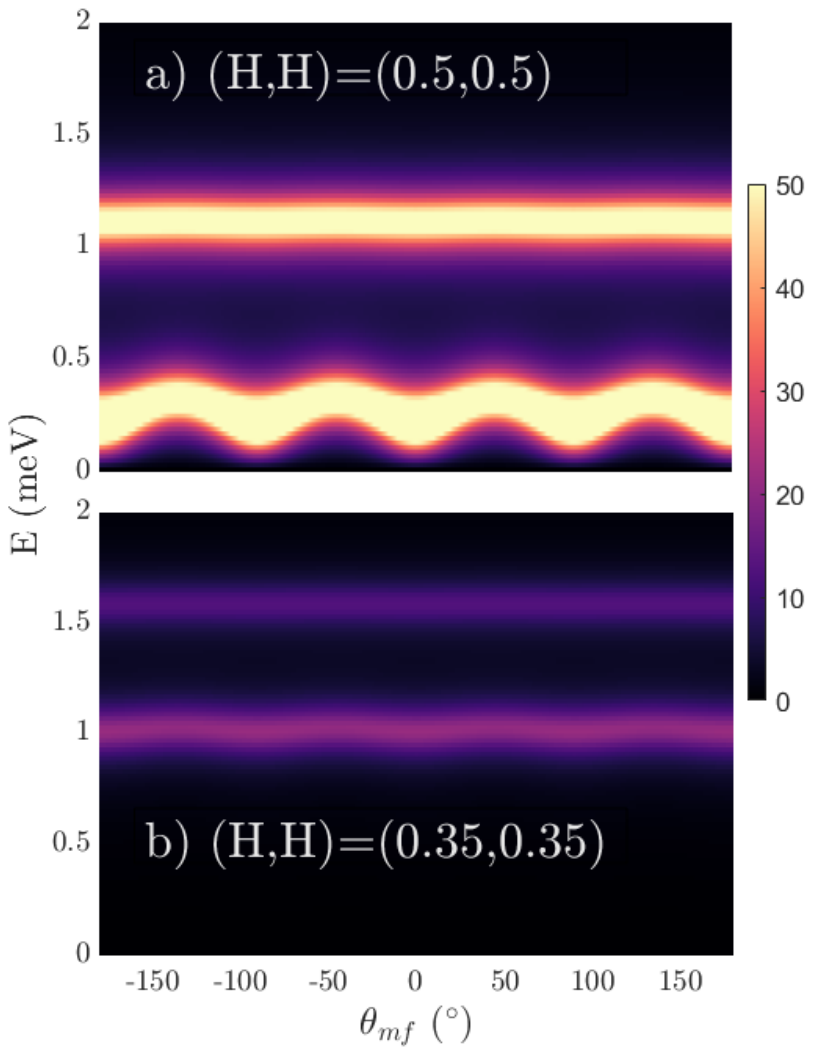}
	\end{center}
	\caption{Plots of the calculated neutron scattering intensity as a function of angle within the $a-b$ plane of the molecular field.  This illustrates the effects of the approximation that the rotation matrix commutes with the single-ion susceptibility ($[\underline{\underline{g}}(\omega),\underline{\underline{R}}_{i}] \equiv 0$) discussed in the main text.}
	\label{fig:angle_check}
\end{figure}

\noindent where $\underline{\underline{\mathcal{J}}}_{mn}$ is the exchange coupling between Ce$^{3+}$ sites at positions ${\bf{r}}_{m}$ and ${\bf{r}}_{n}$. Incorporating the rotation matrices into the definition of $\underline{\underline{\mathcal{J}}}$ in the Dyson equation  allows the exponential factors in the rotation matrices to be incorporated into  $\tilde{J}({\bf{q}})$ as follows, 

\begin{equation}
	\begin{split}
		&\underline{\underline{\mathcal{J}}}({\bf{q}}) \rightarrow 	\underline{\underline{\tilde{\mathcal{J}}}}({\bf{q}})=  \\
		& \qquad \underline{\underline{\mathcal{J}}}({\bf{q}}+{\bf{Q}})\underline{\underline{\Phi}}+\underline{\underline{\mathcal{J}}}({\bf{q}}-{\bf{Q}})\underline{\underline{\Phi}}^{*}+\underline{\underline{\mathcal{J}}}({\bf{q}})\mathbf{n}\mathbf{n}^{T}. \nonumber
	\end{split}
\end{equation}

\noindent The compact form of $\underline{\underline{\tilde{G}}}({\bf{q}},\omega)$ that is now Fourier transformed and written in terms of matrices can be readily solved 

\begin{equation}
	\underline{\underline{\tilde{G}}}({\bf{q}},\omega)=\underline{\underline{g}}(\omega) \left[ \underline{\underline{\mathds{1}}}-\underline{\underline{\tilde{\mathcal{J}}}}({\bf{q}}) \underline{\underline{g}}(\omega) \right]^{-1}.\nonumber
\end{equation}

\noindent Note that this equation is still in the rotating frame.  To transform to the lab frame so that we can obtain the original Green's function from which we can calculate the neutron scattering response, we rotate back as follows

\begin{equation}
	\begin{split}
		\underline{\underline{G}}(\mathbf{q},\omega)=\mathbf{n}\textbf{n}^{T}\underline{\underline{\tilde{G}}}(\mathbf{q},\omega)\mathbf{n}\textbf{n}^{T}\\+\Phi^{*}\underline{\underline{\tilde{G}}}(\mathbf{q}+\mathbf{Q},\omega)\Phi^{T}\\+\Phi\underline{\underline{\tilde{G}}}(\mathbf{q}-\mathbf{Q},\omega)\Phi^{*T}.
	\end{split}
	\nonumber
\end{equation}

\noindent Note that $\Phi^{*T}\equiv \Phi^{\dagger}$, however we have kept complex conjugate transpose for consistency of notation and to avoid confusion between the transpose and the Hermitian conjugate (or adjoint).  This approach differs slightly from our previous work with CoO where we solved a set of linear equations for the various components of $G$ and the incorporation of the rotating frame allows us to combine all of the various sites in the lattice into one single RPA equation for the Green's function response.  Above in the main text, we applied this to the case of  CeRhIn$_{5}$.

\subsubsection{Rotating frame approximations}

An approximation applied above when moving to the rotating frame was that the rotation matrices commuted with the single-ion susceptibility ($[\underline{\underline{g}}(\omega),\underline{\underline{R}}_{i}] \equiv 0$).  Under a rotation transformation, this left the first term in the Dyson expansion unchanged while allowed us to commute the rotation matrix to surround the exchange term $\mathcal{J}$.  It is not clear if this approximation is appropriate given the very anisotropic crystal field terms for Ce$^{3+}$ in CeRhIn$_{5}$ and the complexity of the Stevens' operators required by symmetry.  We also note that unlike transition metal ion counterparts where the crystal field terms are several orders of magnitude larger than exchange energetics, the anisotropic crystal field terms and exchange parameters are of similar sizes in the rare earth $4f$ elements.

We investigate the limits of this approximation in Fig. \ref{fig:angle_check}.  In this calculation we have chosen two momentum positions at the magnetic zone center ${\bf{q}}=(0.5,0.5)$ and away from the zone center near the peak of the dispersion at ${\bf{q}}=(0.35,0.35)$.  In this figure we have rotated the molecular field within the $a-b$ plane (denoted as $\theta_{mf}$).  To give an energetic gap to the lowest energy mode, we have arbitrarily chosen this calculation with a slightly larger $h_{MF}$=0.65 meV over the value of 0.63 meV motivated above.  While there is a small $\sim$0.1 meV variation in the low-energy excitation polarized within the $a-b$ plane, the rotation does not change the number of modes, or the energy scale in comparison to the overall bandwidth of the excitations.  We therefore use this plot as justification for commuting the rotation matrices through the single-ion susceptibility $g(\omega)$ at the mean-field level above in the Dyson equation.


%

\end{document}